\documentclass[reprint, amsmath, amssymb, aps, prd, superscriptaddress, nolongbibliography, nofootinbib]{revtex4-1}

\usepackage{float}
\usepackage{amssymb, amsmath, bm, dcolumn, epsf, graphicx, latexsym, slashed, simplewick}
\usepackage[utf8]{inputenc}
\usepackage{subfigure}
\usepackage{comment}
\usepackage{graphicx}
\usepackage{hyperref}
\usepackage{tabularx}
\usepackage{enumerate}
\usepackage{xspace}
\usepackage{fontawesome}
\usepackage{setspace}
\usepackage{lipsum}

\usepackage[usenames,dvipsnames]{xcolor}
\hypersetup{
    colorlinks = true,
    citecolor = {MidnightBlue},
    linkcolor = {BrickRed},
    urlcolor = {BrickRed}
}

\newcommand{\nmt}{{\tt NaMaster}\xspace}

\newcommand{\bU}{\langle bU\rangle}
\newcommand{\bpe}{\langle bP_e\rangle}
\newcommand{\bsfr}{\langle b\rho_{\rm SFR} \rangle}

\newcommand{\nv}{\hat{\bf n}}

\newcommand{\wisc}{WI$\times$SC\xspace}
\newcommand{\lrg}{DESI LRG\xspace}
\newcommand{\planck}{\emph{Planck}\xspace}
\newcommand{\act}{\emph{ACT}\xspace}

\begin{document}

\title{Joint tomographic measurement of thermal Sunyaev Zeldovich and the cosmic infrared background}

\author{Adrien La Posta}
\email{adrien.laposta@physics.ox.ac.uk}
\affiliation{Department of Physics, University of Oxford, Denys Wilkinson Building, Keble Road, Oxford OX1 3RH, United Kingdom}
\author{David Alonso}
\affiliation{Department of Physics, University of Oxford, Denys Wilkinson Building, Keble Road, Oxford OX1 3RH, United Kingdom}
\author{Carlos Garc\'ia-Garc\'ia}
\affiliation{Department of Physics, University of Oxford, Denys Wilkinson Building, Keble Road, Oxford OX1 3RH, United Kingdom}
\affiliation{Waterloo Centre for Astrophysics, University of Waterloo, Waterloo, ON N2L 3G1, Canada}
\affiliation{Department of Physics and Astronomy, University of Waterloo, Waterloo, ON N2L 3G1, Canada}
\author{Sara Maleubre}
\affiliation{Department of Physics, University of Oxford, Denys Wilkinson Building, Keble Road, Oxford OX1 3RH, United Kingdom}

\date{\today}

\begin{abstract}
  We present a novel method for the tomographic reconstruction of the bias-weighted mean electron pressure $\bpe$ and star formation rate density $\bsfr$, by simultaneously modelling the contribution from the thermal Sunyaev-Zel'dovich (tSZ) effect and the Cosmic Infrared Background (CIB) to the cross-correlation between photometric galaxy samples and multi-frequency Cosmic Microwave Background (CMB) maps. The resulting measurements are independent of the galaxy clustering properties and robust against cross-contamination between tSZ and CIB. Applying this method to publicly available data, we reconstruct the cosmic evolution of $\bpe$ and $\bsfr$ out to $z\sim1$, making our measurements publicly available. Our measurements of both quantities are broadly compatible with predictions from the fiducial FLAMINGO hydrodynamical simulation, although we observe a lower gas pressure at low redshifts, in agreement with other measurements.
\end{abstract}

\maketitle

\paragraph*{\texorpdfstring{\textbf{Introduction.}}{Introduction}}
  The cross-correlation of projected tracers of the large-scale structure (LSS) with the overdensity of galaxies at different redshifts (in combination with the auto-correlation of these galaxies) can be used to reconstruct the cosmic evolution of the astrophysical quantities probed by these tracers, a technique known as ``tomographic reconstruction'' \cite{0805.1409,1302.0857,1810.00885,2006.14650,2504.05384}. Specifically, when applied to cross-correlations with a projected tracer of a field $U$ with galaxies at redshift $z$, tomographic reconstruction is able to recover a measurement of $\bU(z)$; i.e. the mean value of $U$ at that redshift averaged over dark matter haloes, and weighted by the linear halo bias. Recently~\cite{2508.05319} developed and validated a method for tomographic reconstruction that ensures that the recovered values of $\bU$ are completely independent of the clustering properties of the galaxy sample used in the measurement.

  Tomographic reconstruction has been applied to a wide variety of projected probes, leading to the recovery of the bias-weighted mean thermal pressure $\bpe$ from tSZ maps~\cite{1608.04160,1904.13347,1909.09102,2006.14650,2210.08633}, the star-formation rate (SFR) density $\bsfr$ from CIB maps~\cite{2209.05472,2310.10848,2504.05384}, and in general the mean cosmic emissivity at multiple wavelengths \cite{1810.00885,2307.14881,2311.17641}. The method has also been applied to non-electromagnetic probes, leading to bounds on the production rate of gravitational waves \cite{2406.19488}, and high-energy neutrinos \cite{2507.14926} from extragalactic sources.

  The case of $\bpe$ is particularly interesting for cosmology and astrophysics. As these measurements are sensitive to the thermodynamics of the intergalactic medium (IGM) \cite{2007.01679}, they may allow us to discriminate between different baryonic feedback scenarios \cite{2412.12081}, and could also encode useful cosmological information \cite{2105.15043,2309.16323}. However, these measurements are typically complicated by the presence of cross-contamination from the CIB in maps of the tSZ Compton-$y$ parameter reconstructed from multi-frequency observations of the Cosmic Microwave Background (CMB) \cite{1409.6747,2009.05557}. This contamination has been shown to be a major impediment in constraining the thermal properties of gas around dark matter haloes, particularly at relatively high redshifts. Typically this can be diagnosed through the use of so-called ``CIB-deprojected'' tSZ maps, constructed by effectively nulling out the contribution from CIB assuming different infrared spectra (and including potential variability in these spectra) \citep{1006.5599,2307.01043,2307.01258}, although it is not uncommon to find inconclusive results regarding the actual level of contamination \citep{2502.08850}.

  In this paper we present an alternative approach to the tomographic recovery of $\bpe$ in the presence of CIB contamination, based on the separation of both components at the level of the cross-correlation between galaxies and multi-frequency maps. Although a similar approach was presented in \cite{2006.14650}, the analysis presented here differs in a number of key points. First, we apply the method presented in \cite{2508.05319}, guaranteeing that our measurements are independent of the clustering properties of the galaxies used and of their specific infrared spectrum. Secondly, we use an empirical library of templates of infrared spectral energy distributions (SEDs) to model the CIB spectrum as a function of redshift \citep{1304.3936,1409.5796,1703.08795}, rather than assuming an effective modified black-body spectrum. Moreover, we turn the problem of CIB contamination into an opportunity by using our measurements to simultaneously measure $\bpe$ and $\bsfr$ and reconstruct the evolution of two quantities of astrophysical interest. Finally, we apply this method to cross-correlations with high-density photometric galaxy samples, increasing the precision of our measurements compared to those that could be obtained from sparser spectroscopic catalogues. We make these measurements publicly available, together with all the information needed to incorporate them into any cosmological or galaxy evolution study.

    \paragraph*{\texorpdfstring{\textbf{Methods---Tomography.}}{Methods - Tomography}}
    Our measurement of $\bpe$ and $\bsfr$ relies on the methodology developed in \cite{2508.05319} for the extraction of these quantities from the cross-correlations of a tracer of the LSS $U$ with a given galaxy sample, combined with the auto-correlation of these galaxies. As shown in \cite{2508.05319}, the recovered value of $\bU$ is independent of the clustering properties of the sample, and it may be safely interpreted within the halo model as the halo-averaged value of $U$ weighted by halo bias:
    \begin{equation}
      \bU = \int dM\,n(M)\,b_h(M)\,\tilde{U}(M),
    \end{equation}
    where $n(M)$ and $b_h(M)$ are the mass function and halo bias, and $\tilde{U}$ is the volume integral of $U$.

    In this paper we adapt the method presented in \cite{2508.05319} to the use of angular power spectra. Specifically, we model the galaxy-galaxy and galaxy-$U$ correlations as
    \begin{align}
      C_\ell^{gg} &= b_g^2C_\ell^{mm, gg} + A_{gg}C_\ell^{mmk^2,gg} + N_{gg}\label{eq:exp_gg}, \\
      C_\ell^{gU} &= b_g\bU C_\ell^{mm,gU} + A_{gU}C_\ell^{mmk^2,gU} + N_{gU},\label{eq:exp_gU}
    \end{align}
    where
    \begin{align}
      C_\ell^{mm, XY} &= \int \frac{d\chi}{\chi^2}\,q_{X}(\chi)\,q_Y(\chi) \,P_{mm}\left(k, z\right), \\
      C_\ell^{mmk^2, XY} &= \int \frac{d\chi}{\chi^2} q_{X}(\chi)\,q_Y(\chi) \,k^2P_{mm}\left(k, z\right).
    \end{align}
    Here, $P_{mm}(k)$ is the matter power spectrum, and $q_X(\chi)$ is the radial kernel associated with tracers $X$. $b_g$ is the linear galaxy bias of the sample, and $\{A_{gg},N_{gg},A_{gU},N_{gU}\}$ are nuisance parameters that absorb the impact of non-linear, stochastic, and scale-dependent bias. As shown in \cite{2508.05319}, this model is able to recover unbiased measurements of $\bU$ using Fourier scales up to $k_{\rm max}=0.3\,{\rm Mpc}^{-1}$.
  
    Since the model is linear in $A_{XY}$ and $N_{XY}$, we marginalise over these parameters analytically, and only sample $b_g$ and $\bU$. In the following, we focus on $U=P_e,~\rho_{\rm SFR}$ using both the tSZ and CIB signals extracted from CMB multi-frequency observations. The associated radial kernels are
    \begin{align}
      q_{{\rm tSZ},\nu}(\chi) = S_\nu^{\rm tSZ}\,\sigma_T/[m_ec^2(1+z)] \\
      q_{{\rm CIB},\nu}(\chi) = S_\nu^{\rm CIB}(z)\,\chi^2/K,
    \end{align}
    where $\sigma_T$ is the Thomson scattering cross section, $K=10^{-10}~{\rm M_\odot/yr/L_\odot}$ is the calibration between far infrared luminosity and star formation rate~\cite{9807187}, $S^{\rm tSZ}_\nu$ is the universal tSZ spectrum, and $S^{\rm CIB}_\nu(z)$ is the average flux of infrared sources at redshift $z$ normalised to a total luminosity $L_\odot$. Likewise, the galaxy kernel is $q_g(\chi)=H(z)\,p(z)$, where $H$ is the expansion rate, and $p(z)$ is the redshift distribution of the sample. Cross-correlations with galaxies are thus only sensitive to the properties of the tracer $U$ in the redshift range covered by $p(z)$.

  \paragraph*{\texorpdfstring{\textbf{Methods---Multi-frequency correlations.}}{Methods - Multi-frequency correlations}}

    The conventional approach to cross-correlation analyses involving CMB secondary signals is to use component separated maps extracting the component of interest~\cite{1502.01596,2307.01043,2305.10193,2307.01258,2602.11279} to then correlate them with other large scale structure tracers. While it is technically possible to deproject any contaminant from the resulting map with a given spectral energy distribution~\cite{1006.5599,2307.01043}, it comes at the cost of an increased variance in the measured power spectra. Additionally, this deprojection step assumes a known frequency dependence of contaminant signals, which makes the interpretation of residual contamination in the resulting map challenging. Methods have been recently developed to circumvent this issue, following an iterative approach~\citep[e.g.][]{2505.14644}. Instead, here we proceed by modelling the impact of the different components directly at the level of the measured power spectra, similar in spirit to multi-frequency power spectrum-based CMB analyses.

    Assuming that the only components that correlate with the galaxy sample are the tSZ and CIB, we start by modelling the map at frequency $\nu$ as
    \begin{equation}
      m_\nu(\nv) = S_\nu^\mathrm{tSZ}y(\nv) + S_\nu^{\mathrm{CIB}}(z_g)\,c_{z_g}(\nv) + \tilde{N}_\nu(\nv),
    \end{equation}
    where $c_{z_g}(\nv)$ is the contribution to the CIB from the range of redshifts that show significant correlation with the galaxy sample under study, and $y(\nv)$ is the Compton-$y$ map. All other sky components which do not correlate with our target sample, including instrumental noise, the primary CMB, the CIB at other redshifts, etc., are combined into the effective noise term $\tilde{N}_\nu$. We ignore other potentially correlated components, such as the integrated Sachs-Wolfe effect, mostly relevant on larger scales than those studied here, and radio point sources, which are masked and further suppressed at the frequencies we use. Within this model, the cross-power spectrum of this map with the target galaxy sample is
    \begin{equation}
      C_\ell^{g\nu} = S_\nu^\mathrm{tSZ}C_\ell^{gy} + S_\nu^{\mathrm{CIB}}(z_g)\,C_\ell^{gc},\label{eq:multifreq_model}
    \end{equation}
    where $C_\ell^{gy}$($C_\ell^{gc}$) is the cross-correlation between our galaxy sample and the Compton-$y$ map (CIB intensity at the sample's redshift). We then use the model from Eq.~\ref{eq:exp_gU} to describe these cross-correlations with $6$ free amplitude parameters ($A_{XY}$, $N_{XY}$) for each redshift bin in addition to the bias parameters $\{b_g,\bpe,\bsfr\}$. Following a similar rationale, it is also possible to easily model and marginalise over other sources of uncertainty, such as overall map calibration or variations in the spectral energy distributions.

    We also investigated a second approach, in which the $C_\ell^{gy}$ and $C_\ell^{gc}$ bandpowers are first estimated from the multi-frequency cross-spectra, before interpreting the result in terms of the model in Eq. \ref{eq:exp_gU}. Eq. \ref{eq:multifreq_model} may be written as a linear problem of the form ${\bf d}={\sf S}\,{\bf C}+{\bf n}$, where ${\bf d}$ is a vector with all multi-frequency power spectrum measurements, ${\sf S}$ is a matrix containing the tSZ and CIB SEDs as columns, ${\bf C}$ contains the two spectra we wish to recover, and ${\bf n}$ is a noise term. The best-fit estimate of ${\bf C}$ is the well-known least-squares solution: 
    \begin{equation}
      \hat{\bf C} = ({\sf S}^\top\mathsf{\Sigma}^{-1}{\sf S})^{-1}\,{\sf S}^\top\mathsf{\Sigma}^{-1}\,{\bf d},
    \end{equation}
    where $\mathsf{\Sigma}$ is the full multi-frequency power spectrum covariance, and the covariance of $\hat{\bf C}$ is  $({\sf S}^\top\mathsf{\Sigma}^{-1}{\sf S})^{-1}$. As we show in the explanatory material, this approach is mathematically equivalent to the Spectral ILC method of \cite{2509.10604} in the context of cross-correlations. We will refer to this approach as maximum-likelihood bandpower reconstruction (MLBR) in what follows.

    In our analysis we will use the CIB SEDs modelled by~\cite{1304.3936,1409.5796,1703.08795}, and used in past CIB studies \citep{2006.16329,2204.01649,2206.15394}. We will also consider potential variations in the shape of the far infrared spectrum. As a proxy of this effect, we introduce a new parameter $\delta \beta$ for each redshift bin, and rescale the CIB SEDs with a frequency power law $\propto\nu^{\delta\beta}$.

\paragraph*{\texorpdfstring{\textbf{Data.}}{Data}}
  \begin{figure}[t!]
    \centering
    \includegraphics[width=\columnwidth]{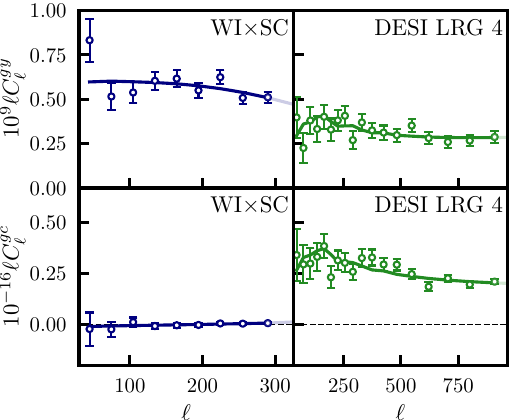}
    \caption{Measurements of the galaxy-tSZ (top row) and galaxy-CIB (bottom row) angular power spectra for the lowest and highest redshift bins used in this analysis. This shows the spILC (or maximum-likelihood) estimates from galaxy-$T^\nu$ measurements. While we have a clear detection of both galaxy-tSZ and galaxy-CIB correlations for the fourth DESI LRG redshift bin, we measure a galaxy-CIB correlation consistent with zero from the \wisc sample.}
    \label{fig:spectra}
  \end{figure}

  We use cross-correlations between multi-frequency maps and galaxies at different redshifts. Specifically, we use the photometric Luminous Red Galaxy (LRG) sample from the DESI Legacy Survey~\cite{2309.06443}, divided into four redshift bins spanning the range $0.4\lesssim z\lesssim 1$. We use the publicly available data and random catalogues, as well as the calibrated redshift distributions for each bin. We complement this with a sample at low redshifts selected from the WISE$\times$SuperCOSMOS photometric galaxy catalog~\cite{1607.01182} (hereafter \wisc). Specifically, we select galaxies in the photometric redshift range $z_p\in[0.15,0.4]$, resulting in a sample covering the low-redshifts $0.1\lesssim z\lesssim 0.4$ missed by the LRG samples. We construct maps of the galaxy overdensity field for each of these samples following the procedures described in \cite{2510.17796,2510.09563}, including the treatment of sky systematics, and calibration of the \wisc redshift distribution.

  As a tracer of extra-Galactic emission, we use temperature maps from the Planck PR4 (NPIPE) data release~\cite{2007.04997} at 100, 143, 217, 353, 545, and 857 GHz. It is worth noting that, while the 100-545 GHz channels were calibrated to the orbital dipole, the 857~GHz map is still using a planet-based calibration. When varied in the analysis, we impose Gaussian priors on calibration coefficients for each frequency channel with a standard deviation set to the calibration uncertainties quoted in~\cite{2007.04997}. To model dust emission, we apply the following colour corrections to the model: 1.076, 1.017, 1.119, 1.097, 1.068 and 0.995 for the 100, 143, 217, 353, 545 and 857~GHz channels~\citep{1309.0382}, which have been computed from the CIB SEDs used in this analysis~\cite{1304.3936,1409.5796,1703.08795}. For some of our additional tests, we also use publicly-available ACT DR6 + Planck Compton-$y$ ILC maps~\cite{2307.01258}, including their CIB-deprojected versions, which can be compared with the method presented here.

\paragraph*{\texorpdfstring{\textbf{Results.}}{Results}}
  \begin{figure*}[t!]
    \centering
    \includegraphics[width=\textwidth]{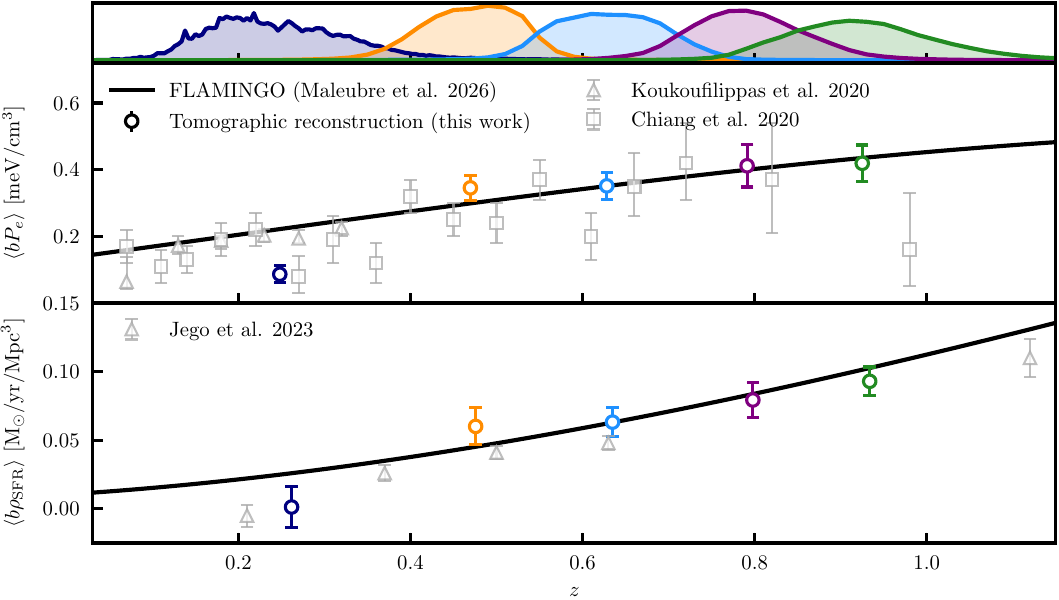}
    \caption{Joint tomographic measurements of bias-weighted electron pressure $\bpe$ and bias-weighted star formation rate density $\bsfr$~from direct correlation of \wisc and \lrg galaxy samples to multi-frequency observations from \planck. We compare to existing measurements~\cite{1909.09102,2006.14650,2206.15394} and to polynomial fits to measurements on FLAMINGO simulations studied in~\cite{2508.05319}.}
    \label{fig:baseline_constraints}
  \end{figure*}
  We applied the tomographic reconstruction method described above to the \wisc and \lrg galaxy samples. We impose a scale cut of $k_{\rm max} = 0.3~{\rm Mpc^{-1}}$ as validated in~\cite{2508.05319}, ensuring that the recovered $\bpe$ and $\bsfr$ are accurate and independent of the galaxy clustering properties a the percent-level. This translates into a maximum multipole $\ell_{\rm max}$ of 308, 581, 749, 851 and 966 for each redshift bin. We use a large-scale cut of $\ell_{\rm min}=30$, which are more sensitive to large-scale systematics. Using the MLBR method, we measure the tSZ-galaxy correlation at high significance in all redshift bins, with a signal-to-noise ratio (SNR) of 37.7, 30.5, 34.4, 30.0 and 30.0, within the scale cuts above, in increasing redshift order. The CIB-galaxy correlation is detected in the LRG redshift bins, with a SNR of 13.9, 21.3, 28.5 and 36.3, whereas the large-scale angular correlation with the low-redshift \wisc sample is consistent with zero (SNR=1.9), in agreement with previous findings~\cite{2206.15394} at similar redshifts, and with theoretical expectations based on the expected star formation history peaking at high redshifts. Figure~\ref{fig:spectra} shows the MLBR estimates of the galaxy-tSZ and galaxy-CIB angular power spectra along with their best-fit model for the lowest and highest tomographic bins. This demonstrates the possibility of a joint extraction of tSZ and CIB cross-correlation signals from multi-frequency power spectra without relying on pre-processed component separated sky maps.

  The values of $\bpe$ and $\bsfr$ reconstructed directly from the multi-frequency cross-correlations are shown in Figure~\ref{fig:baseline_constraints} and summarized in Table~\ref{tab:baseline_constraints}. With some exceptions, discussed below, these results are broadly consistent with previous measurements of $\bpe$~\cite{1909.09102,2006.14650} and $\bsfr$~\cite{2206.15394}, although there are notable differences in those analyses, both in terms of measurement techniques and in the potential impact of small-scale galaxy bias. Our measurements are also in remarkably good agreement with the results from the FLAMINGO hydrodynamical simulation for its fiducial feedback model (shown as solid lines in the figure, as reported in~\cite{2508.05319}). Furthermore, the power spectrum model used to extract these measurements provides an acceptable description of the data, with a probability-to-exceed (PTE) of 11.8\%.

  Our measurement of $\bsfr$ in the \wisc redshift bin lies marginally below the FLAMINGO prediction, and is compatible with zero. This is in agreement with earlier measurements from \cite{2206.15394} at similar redshifts using a different sample. We tested for the impact of potential evolution in $\rho_{\rm SFR}$ within the \wisc redshift distribution, allowing for a $\propto z^2$ behaviour at low redshift, and found no significant differences in the recovered constraints. We also find a perhaps surprisingly low value of $\bpe$ in this first redshift bin, although the measurement is in agreement with previous analyses at the same redshift \citep{2006.14650}. We tested for the impact of sky contamination in the observed galaxy number counts, manifesting itself as extra large-scale power in the galaxy auto-spectrum, and thus leading to a downwards bias in the inferred $\bpe$. Specifically, we repeated our analysis using a more conservative Galactic mask ($f_{\rm sky} = 0.4$) to reduce the impact of dust contamination, and considered using a more conservative large-scale cut $\ell_{\rm min}=50$ and $80$. In all cases, the value of $\bpe$ recovered in the lowest redshift bin remained consistent with our fiducial analysis within statistical uncertainties. The same applies to $\bsfr$, although we observed a somewhat larger shift of $2.4\sigma$ when using the smaller Galactic mask for the first \lrg bin.

\begin{figure}[t!]
    \centering
    \includegraphics[width=\columnwidth]{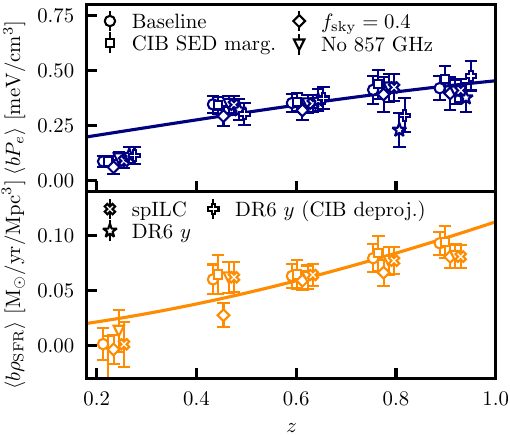}
    \caption{$\bpe$ and $\bsfr$ measurements derived from different analysis settings. We included a comparison with ACT DR6 Compton-$y$ maps~\cite{2401.13033}. Solid lines show $\bpe$ and $\bsfr$ measurements on the FLAMINGO simulation for its fiducial feedback model from~\cite{2508.05319}.}
    \label{fig:robustness}
  \end{figure}

  To quantify the potential impact of deviations with respect to the model infrared SEDs of~\cite{1304.3936,1409.5796,1703.08795} used here, we introduced an additional parameter $\delta \beta$, modifying the tilt of the spectrum as $S_\nu^{\rm CIB}\propto\nu^{\delta\beta}\bar{S}_\nu^{\rm CIB}$, where $\bar{S}^{\rm CIB}_\nu$ is our fiducial CIB SED. Note that the effective SED needs to be renormalised to unit luminosity after this rescaling to avoid biasing the recovered value of $\bsfr$. We introduce one such $\delta\beta$ for each redshift bin, and marginalise over these parameters assuming a flat prior in the range $\delta\beta\in[-2,2]$. We find that marginalising over this parameter only leads to a mild increase in the statistical uncertainties of the recovered $\bpe$ and $\bsfr$, with no significant shift in their best-fit values. We observe a mild improvement in the goodness-of-fit, with a PTE of 16.5\%. We find no evidence of a departure from our fiducial CIB SEDs, and the data is able to constrain such departures with a precision of $\sigma(\delta\beta)\sim0.1$ in all but the lowest redshift bin (where the CIB signal itself is compatible with zero). Similarly, marginalizing over CMB map calibrations from~\cite{2007.04997} or discarding some frequency bands from the analysis do not change or degrade our measurements significantly. A summary of these tests is shown in Figure \ref{fig:robustness}.

  \begin{table}[t!]
    \centering
    \begin{ruledtabular}
    \begin{tabular}{lllcc}
        \multicolumn{1}{c}{} & \multicolumn{1}{c}{} & \multicolumn{1}{c}{} & \multicolumn{1}{c}{${\boldsymbol{\langle bP_e \rangle}}$} & \multicolumn{1}{c}{$\boldsymbol{\langle b\rho_{\rm SFR} \rangle}$}\\
        \multicolumn{1}{c}{${\boldsymbol{\bar{z}_{\rm clust}}}$} & \multicolumn{1}{c}{$\boldsymbol{\bar{z}_{\rm CIB}}$} & \multicolumn{1}{c}{$\boldsymbol{\bar{z}_{\rm tSZ}}$} &  [${\mathrm{meV}.\mathrm{cm}^{-3}}$] & [$\mathrm{M_\odot}.\mathrm{yr}^{-1}.\mathrm{Mpc}^{-3}$] \\[3.0pt]
        \hline
        \noalign{\vskip 3.0pt}
        0.254 & 0.262 & 0.248 & $0.086~\pm~0.026$ & $0.001~\pm~0.015$ \\
        0.472 & 0.476 & 0.470 & $0.345~\pm~0.037$ & $0.060~\pm~0.014$ \\
        0.631 & 0.635 & 0.628 & $0.351~\pm~0.040$ & $0.063~\pm~0.011$ \\
        0.795 & 0.798 & 0.791 & $0.412~\pm~0.063$ & $0.079~\pm~0.013$ \\
        0.930 & 0.934 & 0.925 & $0.419~\pm~0.055$ & $0.093~\pm~0.010$ \\
    \end{tabular}
    \end{ruledtabular}
    \caption{Joint tomographic measurements of bias-weighted electron pressure $\bpe$ and bias-weighted star formation rate density $\bsfr$ as shown in Figure~\ref{fig:baseline_constraints}.}
    \label{tab:baseline_constraints}
  \end{table}

  We also applied the tomographic reconstruction techniques to the component-separated power spectra obtained using the MLBR approach, finding results compatible with our fiducial analysis. Finally, we applied our method to cross-correlations between our galaxy samples and publicly available \planck+\act Compton-$y$ maps~\cite{2307.01258}, including an example of a CIB deprojected map ($\beta=1.4$, $T_d=24~\rm{K}$). The resulting measurements, shown in Fig. \ref{fig:robustness}, are broadly consistent with our baseline constraints, although we observe a more significant downwards shift in the third LRG bin ($z\sim0.8$). This is not entirely surprising, as some level of CIB contamination is expected, particularly at higher redshifts, which would reduce the cross-correlation amplitude.

\paragraph*{\texorpdfstring{\textbf{Conclusion.}}{Conclusion}}
  We have presented a method to recover tomographic measurements of the bias-weighted mean pressure $\bpe$ and the star formation-rate density $\bsfr$ from the cross-correlation between multi-frequency temperature maps and samples of galaxies at different redshifts. The method draws from similar approaches previously used in the literature \cite{1810.00885,2006.14650,2504.05384}, combining them with a robust model~\cite{2508.05319} guaranteeing that the resulting measurements are fully independent from the small-scale galaxy bias. The method also improves on previous studies of the intergalactic thermal gas pressure based on the use of component-separated tSZ maps, turning one of their main weaknesses -- CIB contamination -- into an opportunity to recover precise and complementary measurements of the cosmic star formation history.

  Applying this method to \planck data and galaxies from the \wisc survey and the DESI Legacy Survey, we have made measurements of $\bpe$ and $\bsfr$ over the redshift range $z\lesssim1$, and present them here with all the information necessary to incorporate them as external datasets in analyses targetting the thermal properties of the IGM, galaxy formation, and cosmology. We have shown that our measurements are robust to potential systematics, including the mis-modelling of the CIB spectrum. They are also broadly compatible with standard predictions from hydrodynamical simulations, such as FLAMINGO, although we recover a lower gas pressure measurement at the lowest redshift probed ($z\sim0.25$). 

  The methodology presented here could be extended to study the small-scale tSZ and CIB signature in cross-correlations with galaxies and other LSS tracers, such as cosmic shear. A key challenge in this case is the potentially different spectral properties of the CIB on large-scales (dominated by the collective emission of all infrared-emitting sources), and on small-scales (i.e. in the one-halo regime), where the specific tracer used determines the spectral behaviour of the cross-correlation. This is further complicated by the need to account for the small-scale clustering properties of galaxies. Following the rationale presented here, to obtain robust constraints on the small-scale properties of the IGM, these effects must be forward-modelled to the space of multi-frequency cross-correlations, rather than relying solely of component-separated maps.

\section{Acknowledgments}
  We would like to thank Matthieu Bethermin and Baptiste Jego for useful discussions regarding the infrared spectral templates used here, and for making them publicly available. ALP and DA are supported by the Science and Technology Facilities Council (STFC) under grants with reference UKRI1164 and ST/W000903/1. ALP, DA, SM, and CGG acknowledge support from the Beecroft Trust.

\bibliography{draft}

\appendix

\onecolumngrid
\newpage


\begin{center}
    {\large \textbf{Explanatory supplement --- Joint tomographic measurement of thermal Sunyaev Zeldovich and the cosmic infrared background}}\\[1.5em]
\end{center}

\section{Methods}\label{sec:meth}
  \subsection{Power spectrum estimation}\label{app:cls}
    All power spectra used here were measured using he catalog-to-cosomlogy pipeline {\tt Cosmotheka}~\cite{cosmotheka, cosmotheka_github}, that uses the pseudo-$C_\ell$ estimator as implemented in \nmt \citep{1809.09603}. The Gaussian covariance matrix of these measurements was estimated analytically using the improved narrow kernel approximation (iNKA) described in \cite{1906.11765,2010.09717}. We used the pseudo-$C_\ell$s measured directly from the data as input for the iNKA.

    When analysing the LRG data we used the publicly available random catalogues\footnote{The DESI LRG catalogs and randoms can be found in \url{https://data.desi.lbl.gov/public/papers/c3/lrg_xcorr_2023/v1/}.} to determine the spatially-varying completeness of the sample. The masked galaxy overdensity field in pixel $p$ was constructed as
    \begin{equation}
      \tilde{\delta}_{g}(\nv_p)=n^g_p-\alpha\,n^r_p,\hspace{12pt}\alpha\equiv\frac{N_g}{N_r},
    \end{equation}
    where $n^g_p$ and $n^r_p$ are the number of galaxies and random points in the pixel, and $N_g$ ($N_r$) is the total number of galaxies (randoms)~\cite{lrg_notebook}. The mask is then given by the expected number of objects in each pixel, determined from the random catalogue: $w_g(\nv_p)=\alpha\,n^r_p$. A similar procedure was used in the case of \wisc, although in this case the mask is simply binary (see \cite{1909.09102,2510.09563} for further details).

    In our fiducial analysis, when analysing the \planck maps, we used the publicly available Galactic mask covering 60\% of the sky. We also considered the result of using the more conservative 40\% mask in one of our consistency tests. In addition to this, we used the HFI point source mask, with the final mask resulting from the product of the Galactic and point-source masks.

    All sky maps were generated using the HEALPix pixelisation scheme \cite{astro-ph/0409513} with resolution parameter $N_{\rm side}=1024$, corresponding to a pixel size of $\sim3.4'$. We computed all power spectra in bandpowers of width $\Delta \ell=30$. When analysing the data for a given redshift bin with mean redshift $\bar{z}$, we discarded all bandpowers with mean $\ell$ larger than $\ell_{\rm max}=k_{\rm max}\chi(\bar{z})$, where $k_{\rm max}=0.3\,{\rm Mpc}^{-1}$ (as validated in \cite{2508.05319}), and $\chi(\bar{z})$ is the comoving distance to redshift $\bar{z}$. We additionally removed all bandpowers with $\ell<30$ to minimise the impact of correlated large-scale systematic contamination in the galaxy and \planck maps. Additionally, we removed all multipoles above $\ell=300$ for correlations involving the 857\,{\rm GHz} map to match existing \planck tSZ map-based reconstructions~\cite{1502.01596,2305.10193}, excluding small scales to mitigate infrared sources contamination in Compton-$y$ maps. Calibration uncertainties are also larger for this frequency channel~\cite{2007.04997}, which we can marginalise over in our approach but is technically more difficult with map-based methods. 

  \subsection{Direct modelling of LSS correlations with CMB multi-frequency maps}\label{app:MFLSS}
    In our fiducial analysis, we recover measurements of our parameters of interest, $b_g$, $\bpe$, and $\bsfr$, by directly modelling the cross-correlation between galaxies and temperature maps at different frequencies, and the auto-correlation of these galaxies. The model described in the main text also includes a number of nuisance parameters needed to account for scale-dependent and stochastic bias in the different tracers of the matter fluctuations explored here (galaxies, gas pressure, and SFR density), $A_{XY}$ and $N_{XY}$. For a given galaxy sample $i$, the full set of parameters is then
    \begin{equation}
      \boldsymbol{\theta} = \left(b_{g_i},\,\bpe_i,\,\bsfr_i,\,A_{g_ig_i},\,A_{g_i P_e},\,A_{g_i\rho_{\rm SFR}},\,N_{g_ig_i},\,N_{g_i P_e}~N_{g_i\rho_{\rm SFR}}\right)^\top
    \end{equation}

    To place constraints on these parameters, we use a Gaussian likelihood of the form:
    \begin{equation}
      -2\log p({\bf d}|\boldsymbol{\theta})=({\bf d}-{\bf t}(\boldsymbol{\theta}))^\top\mathsf{\Sigma}^{-1}({\bf d}-{\bf t}(\boldsymbol{\theta}))+L,
    \end{equation}
    where ${\bf d}\equiv\left(\hat{C}^{g_ig_i}_\ell,\hat{C}^{g_i\,\nu_1},\cdots,\,\hat{C}^{g_i\,\nu_N}_\ell\right)$ are the measurements of the galaxy auto-correlation and all the cross-correlations with $N$ frequency maps, ${\bf t}(\theta)$ is the theoretical prediction, $\mathsf{\Sigma}$ is the covariance matrix of the measurements, and $L$ is a parameter-independent normalisation constant. Although the number of free parameters in our model is relatively large (9 per redshift bin), we can write $t(\boldsymbol{\theta})$ as a linear model of the form:
    \begin{equation}
      {\bf t}_\ell(\boldsymbol{\theta})=\mathsf{C}_\ell\cdot\tilde{\boldsymbol{\theta}},
    \end{equation}
    where $\tilde{\boldsymbol{\theta}}$ is a reparametrisation of $\boldsymbol{\theta}$ that transforms the non-linear parameters $(b_g,\bpe,\bsfr)$ into linear amplitudes:
    \begin{figure*}[t!]
        \centering
        \includegraphics[width=\textwidth]{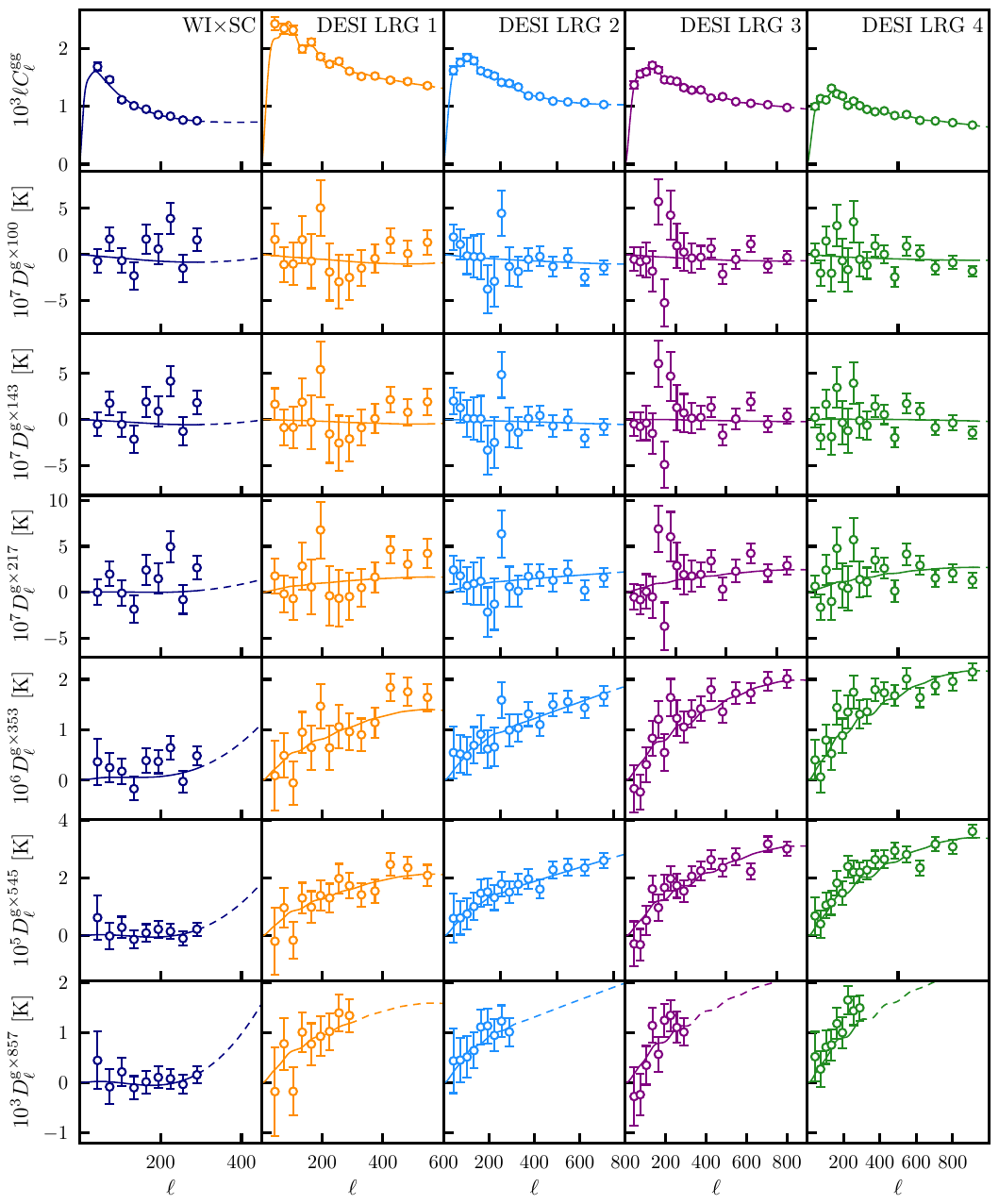}
        \caption{Power spectra measured for \wisc and \lrg auto-correlations (top row) and their correlation with \planck PR4 multi-frequency maps at 100, 143, 217, 353, 545 and 857 GHz. We only show data points with $\ell < k_{\rm max}\chi(\bar{z})$, and $k_{\rm max}<0.3~\rm Mpc$. For correlations involving \planck at 857 GHz, we apply an additional cut at $\ell=300$. }
        \label{fig:cls_array}
    \end{figure*}
    \begin{equation}
      \tilde{\boldsymbol{\theta}} = \left(b_{g_i}^2,\,b_{g_i}\bpe_i,\,b_{g_i}\bsfr_i,\,A_{g_ig_i},\,A_{g_i P_e},\,A_{g_i\rho_{\rm SFR}},\,N_{g_ig_i},\,N_{g_i P_e}~N_{g_i\rho_{\rm SFR}}\right)^\top.
    \end{equation}
    In turn, ${\sf C}_\ell$ is a matrix of power spectrum templates:
    \begin{equation}
      \mathsf{C}_\ell = \begin{pmatrix}
        C_\ell^{mm,g_i\times g_i} & 
        0 &
        0 &
        C_\ell^{mmk^2,g_i\times g_i} &
        0 &
        0 &
        1 & 0 & 0 \\
        0 &
        C_\ell^{mm,g_i\times \nu_1^{\rm tSZ}} & 
        C_\ell^{mm,g_i\times \nu_1^{\rm CIB}} & 
        0 &
        C_\ell^{mmk^2,g_i\times \nu_1^{\rm tSZ}} & 
        C_\ell^{mmk^2,g_i\times \nu_1^{\rm CIB}} & 
        0 &
        S_{\nu_1}^{\rm tSZ} &
        \bar{S}_{\nu_1}^{{\rm CIB},g_i}  \\
        \vdots & \vdots & \vdots & \vdots & \vdots & \vdots & \vdots & \vdots & \vdots \\
        0 &
        C_\ell^{mm,g_i\times \nu_N^{\rm tSZ}} & 
        C_\ell^{mm,g_i\times \nu_N^{\rm CIB}} & 
        0 &
        C_\ell^{mmk^2,g_i\times \nu_N^{\rm tSZ}} & 
        C_\ell^{mmk^2,g_i\times \nu_N^{\rm CIB}} & 
        0 &
        S_{\nu_N}^{\rm tSZ} &
        \bar{S}_{\nu_N}^{{\rm CIB},g_i}  \\
      \end{pmatrix},
    \end{equation}
    where the power spectrum templates are
    \begin{align}
      C_\ell^{mm, XY} &= \int \frac{d\chi}{\chi^2}\,q_{X}(\chi)\,q_Y(\chi) \,P_{mm}\left(k_\ell, z\right), \\
      C_\ell^{mmk^2, XY} &= \int \frac{d\chi}{\chi^2} q_{X}(\chi)\,q_Y(\chi) \,k^2_\ell\,P_{mm}\left(k_\ell, z\right),
    \end{align}
    with $P_{mm}(k,z)$ the matter power spectrum, and $k_\ell\equiv(\ell+1/2)/\chi$. $S_\nu^{\rm tSZ}$ is the tSZ spectrum at frequency $\nu$, and $\bar{S}^{{\rm CIB},g_i}_\nu$ is the CIB spectrum averaged over the redshift distribution of the $i$-th galaxy sample, $p_i$:
    \begin{equation}
      \bar{S}^{{\rm CIB},g_i}_\nu\equiv\int dz\,p(z)\,S^{\rm CIB}_\nu(z).
    \end{equation}
    Finally, the radial kernels are:
    \begin{align}
      q_{\nu_j^{\rm tSZ}}(\chi) = \sigma_T S_{\nu_j}^{\rm tSZ} / [m_ec^2(1+z)],\hspace{12pt}
      q_{\nu_j^{\rm CIB}}(\chi) = \chi^2 S_{\nu_j}^{\rm CIB}(z)/K, \hspace{12pt} q_{g_i}(\chi) &= H(z)~p_i(z)
    \end{align}
    where $H$ is the expansion rate, $p_i(z)$ is the redshift distribution of the $i$-th galaxy sample,  and $K$ is the constant relating infrared luminosity and star formation rate \cite{9807187}. Note that, in this parametrisation, the CIB and tSZ SEDs have been absorbed into the radial kernels entering the power spectrum templates. Since the model is linear in all amplitude-like parameters $\tilde{\boldsymbol{\theta}}$, we follow~\cite{2301.11895} and analytically marginalise over a subset of them ($A_{XY}$ and $N_{XY}$) to restrict the number of sampled parameters to 3 per redshift bin: $\{b_{g_i},\,\bpe_i,\,\bsfr_i\}$. We assumed the best-fit \planck+ ACT cosmological parameters~\cite{2503.14452} to construct the power spectrum template matrix ${\sf C}_\ell$.
    
    This multi-frequency model is flexible enough to marginalise over additional parameters, such as frequency map calibration amplitudes $c_{\nu_j}$ or changes in the spectral tilt of the CIB SED, parametrised via $\delta\beta$ as described in the main text. These can be incorporated in the calculation of the power spectrum template matrix $\mathsf{C}_\ell$ and marginalized over when sampling posterior distributions.

    \begin{figure*}[t!]
        \centering
        \includegraphics[width=\textwidth]{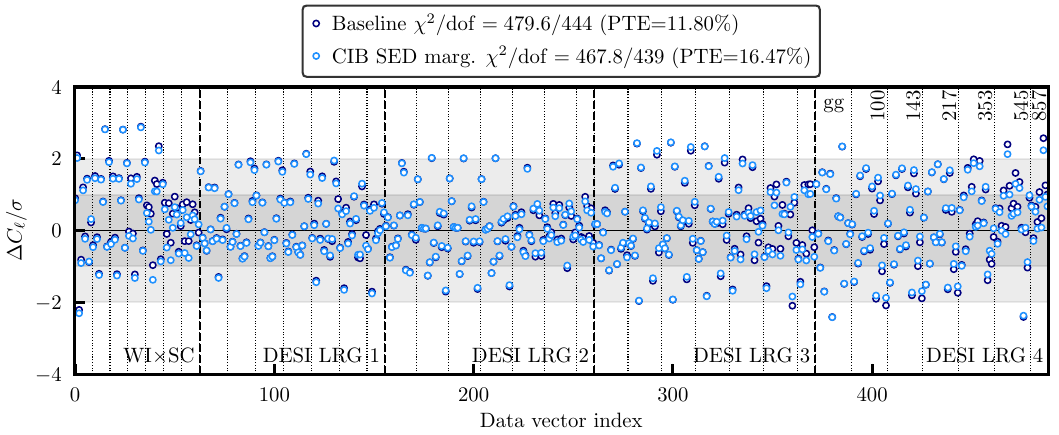}
        \caption{Residual power spectrum vector normalized by the standard deviation. Each block, separated by dashed vertical lines, contains to all bandpowers (auto- and cross-correlations) involving one of the galaxy samples, which we indicate at the bottom of the plot. We show residuals for our fiducial analysis (dark blue) and when marginalising over CIB SEDs (blue). The latter case introduces 5 additional parameters (spectral tilt), improving the goodness-of-fit.}
        \label{fig:cls_residual}
    \end{figure*}

    Our baseline tomographic reconstruction of $\bpe$ and $\bsfr$, discussed in the main text, uses the multi-frequency approach described above. Figure~\ref{fig:cls_array} shows the measured bandpowers for galaxy-temperature correlations for each frequency and redshift bin. As pointed out in the main text, the CIB SED templates assumed provide a reasonable fit to the data with an overall PTE of $11.8\%$. The residual plot in Fig.~\ref{fig:cls_residual} shows a high correlation in the model residuals across the CMB dominated channels (100, 143 and 217~GHz), showing that the uncorrelated noise-like component in these maps is dominated by the CMB primary anisotropies. Taking advantage of the flexibility of our direct multi-frequency modelling method, when marginalizing over the shape of CIB SED templates with a spectral tilt $\delta\beta$, we observe a small improvement in the residuals with a PTE of $16.5\%$ (see cyan points in Fig.~\ref{fig:cls_residual}). Additionally, excluding correlations with the \wisc sample, our fiducial analysis provides an improved fit to the data with a PTE of $20.6\%$.

  \subsection{Maximum likelihood bandpower reconstruction and spectral ILC}\label{app:spILC}
    To test the stability of our results with respect to the details of the forward model used to extract measurements of $\bpe$ and $\bsfr$ directly from multi-frequency cross-correlations, we explored a second approach in which we first extract measurements of the galaxy-tSZ $C_\ell^{gy}$ and galaxy-CIB spectra ($C_\ell^{gy}$ and $C_\ell^{gc}$), and then infer the parameters of interest from these. The method used here is based on a maximum-likelihood estimate of these spectra from the multi-frequency data given a model for the component SEDs. Here we describe the details of the method and show that it is equivalent to the spILC method of~\cite{2509.10604}, a power spectrum-based variant of standard map-based internal linear combination (ILC) approach, when applied to multi-frequency cross-correlations.

    Focusing on a single galaxy redshift bin $g$, we can express a noisy multi-frequency map as
    \begin{equation}
      m_\nu(\nv) = S_\nu^a a(\nv) + S_\nu^b\,b(\nv) + n_\nu(\nv)
    \end{equation}
    where $S_\nu^{a/b}$ are the SEDs of components $a$ and $b$, where $a(\nv)$ is the component of interest and $b(\nv)$ is a contaminant we wish to deproject. Both components exhibit correlation with the target galaxy sample $g$, while $n_\nu(\nv)$ incorporates all map sources uncorrelated with $g$. This includes instrumental noise, Galactic foregrounds, and contributions to $a$ and $b$ from redshifts outside the range covered by $g$.
        
    \begin{figure*}
      \centering
      \includegraphics[width=\textwidth]{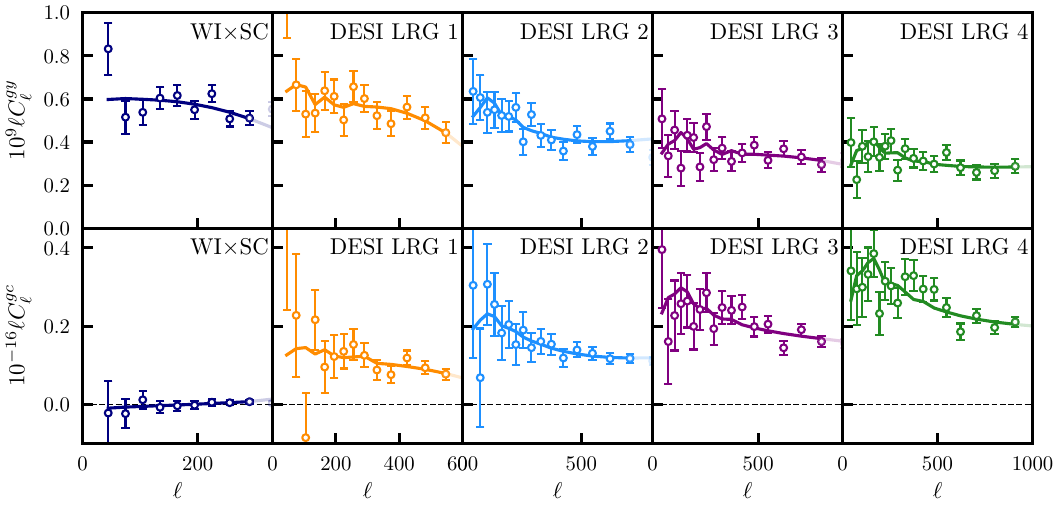}
      \caption{Maximum likelihood bandpower reconstruction for \wisc and \lrg cross-correlations with \planck PR4 multi-frequency maps. We only show data points with $k_{\rm max}<0.3~{\rm Mpc}$. The first panel shows that the galaxy-CIB correlation measured with the \wisc galaxy sample is consistent with zero.}
      \label{fig:MLBRspectra}
    \end{figure*}

    Following the spILC formalism~\cite{2509.10604}, we find optimal weights $W_\ell^\nu$ such that
    \begin{equation}
      \hat{C}^{ga,{\rm ILC}}_\ell\equiv\sum_\nu W_\ell^\nu \hat{C}^{g\nu}_\ell
    \end{equation}
    is an unbiased estimator of the cross-correlation between $a$ and $g$ with minimum variance and zero contribution from component $b$. Here $\hat{C}^{g\nu}_\ell$ is the measured cross-correlation between the target galaxy sample and frequency map $m_\nu$. Writing $W_\ell^{\nu}$, and $S_\nu^{a/b}$ as vectors ${\bf W}_\ell,\,{\bf S}^{a/b}$, the ILC conditions translate into the following equations:
    \begin{align}
      2\mathsf{\Sigma}_\ell\,{\bf W}_\ell-\lambda{\bf S}^a-\mu{\bf S}^b=0,\hspace{12pt}
      {\bf S}^{a\top}{\bf W}_\ell=1,\hspace{12pt}{\bf S}^{b\top}{\bf W}_\ell=0,
    \end{align}
    where $\lambda$ and $\mu$ are two Lagrange multipliers, and $\mathsf{\Sigma}$ is the multi-frequency power spectrum covariance:
    \begin{equation}
      \Sigma^{\nu\nu'}_\ell\equiv{\rm Cov}\left(\hat{C}^{g\nu}_\ell,\hat{C}^{g\nu'}_\ell\right).
    \end{equation}
    Solving for ${\bf W}_\ell$, $\lambda$, and $\mu$, we find the optimal weights
    \begin{equation}\label{eq:weights_spilc}
      \mathbf{W}_\ell = \frac{Q^{bb}_\ell\mathsf{\Sigma}_\ell^{-1}{\bf S}^a-Q^{ab}_\ell\mathsf{\Sigma}_\ell^{-1}{\bf S}^b}{Q^{aa}_\ell Q^{bb}_\ell-(Q^{ab}_\ell)^2},
    \end{equation}
    where $Q^{xy}_\ell\equiv {\bf S}^{x\top}\mathsf{\Sigma}^{-1}_\ell{\bf S}^y$.

    In our approach (labelled maximum-likelihood bandpower reconstruction -- MLBR -- in the main text), we start from a model of the measured multi-frequency correlations as:
    \begin{equation}
      \hat{C}_\ell^{g\nu} = C_\ell^{ga}S_\nu^a + C_\ell^{gb}S_\nu^b+\hat{N}^\nu_\ell,
    \end{equation}
    where the noise component $\hat{N}^\nu_\ell$ includes all chance correlations between $\delta_g$ and $n_\nu$, as well as cosmic variance. Writing the multi-frequency measurements as a vector $\hat{\bf d}\equiv(\hat{C}^{g\nu_1}_{\ell},\cdots,\hat{C}^{g\nu_N}_{\ell})^\top$, the target spectra as ${\bf C}\equiv(C^{ga}_{\ell},C^{gb}_{\ell})^\top$, and the SEDs as a matrix ${\sf S}\equiv({\bf S}^a,{\bf S}^b)$, the model above may be written as:
    \begin{equation}
      \hat{\bf d}={\sf S}\,{\bf C}+\hat{N},
    \end{equation}
    where $\hat{\bf N}$ incorporates all elements of $\hat{N}^\nu_\ell$. Assuming that $\hat{N}$ is Gaussianly distributed, the maximum-likelihood estimator for ${\bf C}$ can be found via the standard least-squares solution:
    \begin{equation}
      {\bf C}_{\rm ML}=\mathsf{\Sigma}_c\,{\sf S}^\top\,\mathsf{\Sigma}^{-1}\,\hat{\bf d},\hspace{12pt}\mathsf{\Sigma}_c\equiv\left({\sf S}^\top\,\mathsf{\Sigma}^{-1}{\sf S}\right)^{-1},
    \end{equation}
    where $\mathsf{\Sigma}_c$ is the covariance matrix of ${\bf C}_{\rm ML}$. Writing this solution in terms of the individual spectra ${\bf S}^{a/b}$, we find:
    \begin{equation}
      \mathbf{C}^{\rm ML} = \frac{1}{\Delta}\begin{pmatrix}
        (\mathbf{S}^{b\top}\mathsf{\Sigma}^{-1}\mathbf{S}^b)\, \mathbf{S}^{a\top}\mathsf{\Sigma}^{-1}\hat{\bf d} - (\mathbf{S}^{a\top}\mathsf{\Sigma}^{-1}\mathbf{S}^b) \mathbf{S}^{b\top}\mathsf{\Sigma}^{-1}\hat{\bf d} \\
        (\mathbf{S}^{a\top}\mathsf{\Sigma}^{-1}\mathbf{S}^a)\mathbf{S}^{b\top}\mathsf{\Sigma}^{-1}\hat{\bf d} - (\mathbf{S}^{b\top}\mathsf{\Sigma}^{-1}\mathbf{S}^a)\mathbf{S}^{a\top}\mathsf{\Sigma}^{-1}\hat{\bf d}
      \end{pmatrix},
    \end{equation}
    where $\Delta\equiv(\mathbf{S}^{a\top}\mathsf{\Sigma}^{-1}\mathbf{S}^a)(\mathbf{S}^{b\top}\mathsf{\Sigma}^{-1}\mathbf{S}^b) - (\mathbf{S}^{b\top}\mathsf{\Sigma}^{-1}\mathbf{S}^a)^2$. This is equivalent to the spILC solution for $C^{ga}_\ell$ with the weights in Eq. \ref{eq:weights_spilc}.

    Once $C_\ell^{g,a/b}$ are measured, we follow the approach from Section~\ref{app:MFLSS} to reconstruct $b_g$, $\bpe$ and $\bsfr$, without the need to model the full multi-frequency data vector, and with the advantage of obtaining model-independent measurements of the component-level cross-correlations as intermediate products. We show these reconstructed bandpowers in Fig.~\ref{fig:MLBRspectra} for the \wisc and \lrg redshift bins. As discussed in the main text, we detect these correlations at high significance in most redshift bins, except for the galaxy-CIB correlation measured with the \wisc galaxy sample, which is consistent with zero.

\begin{figure*}[t!]
    \centering
    \subfigure[~]{\includegraphics[width=0.499\textwidth]{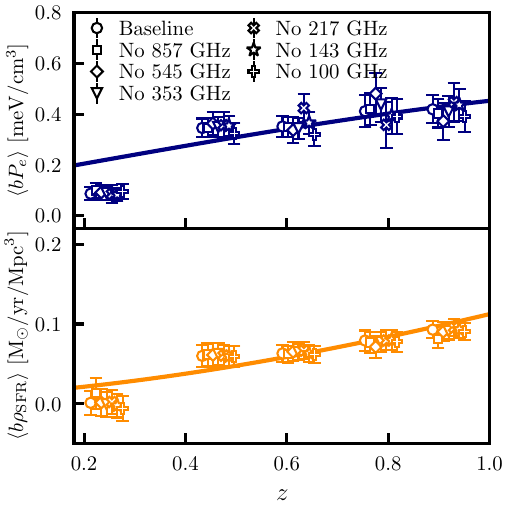}}\hfill
    \subfigure[~]{\includegraphics[width=0.499\textwidth]{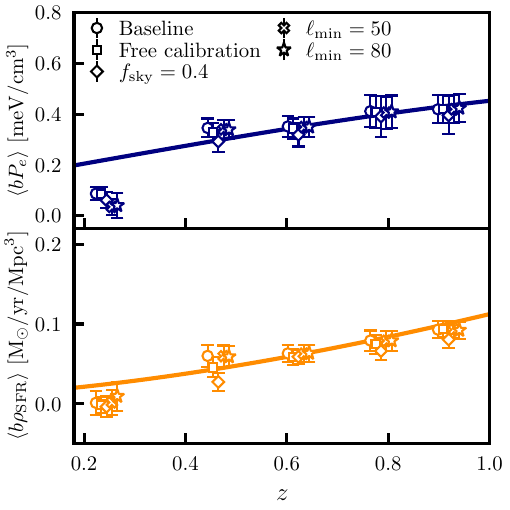}}
    \caption{(a) Multi-frequency reconstructions of $\bpe$ and $\bsfr$ showing the effect of excluding \planck frequency channels from the analysis. (b) We show the measurement stability when varying \planck calibration amplitudes, using a more restrictive Galactic mask or excluding the largest scales from the analysis.}
    \label{fig:robustness_exp}
\end{figure*}

\section{Results}\label{app:res}
  \subsection{Robustness tests}\label{app:res.robustness}
    We have run a number of tests to quantify the robustness of our results against different analysis choices and potential systematics. We present these here. The results of these tests are visually presented in Fig.~\ref{fig:robustness_exp}.
    
    First, we repeated our analysis excluding individual \planck frequencies from the multi-frequency reconstruction of $\bpe$ and $\bsfr$. This allows us to test for potential systematic contamination in any of the frequency maps. The results are shown on the left panel of Fig. \ref{fig:robustness_exp}, and show that our constraints are robust and not driven by any particular frequency channel. We observe the largest variation ($1.8\sigma$) for $\bpe$ at intermediate redshifts ($z\sim0.6$) when removing the 217 GHz channel. This channel is of particular importance, as the tSZ spectrum crosses zero at that frequency. Nevertheless, we do not observe similar shifts at other redshifts.
    \begin{figure*}
      \centering
      \includegraphics[width=0.7\textwidth]{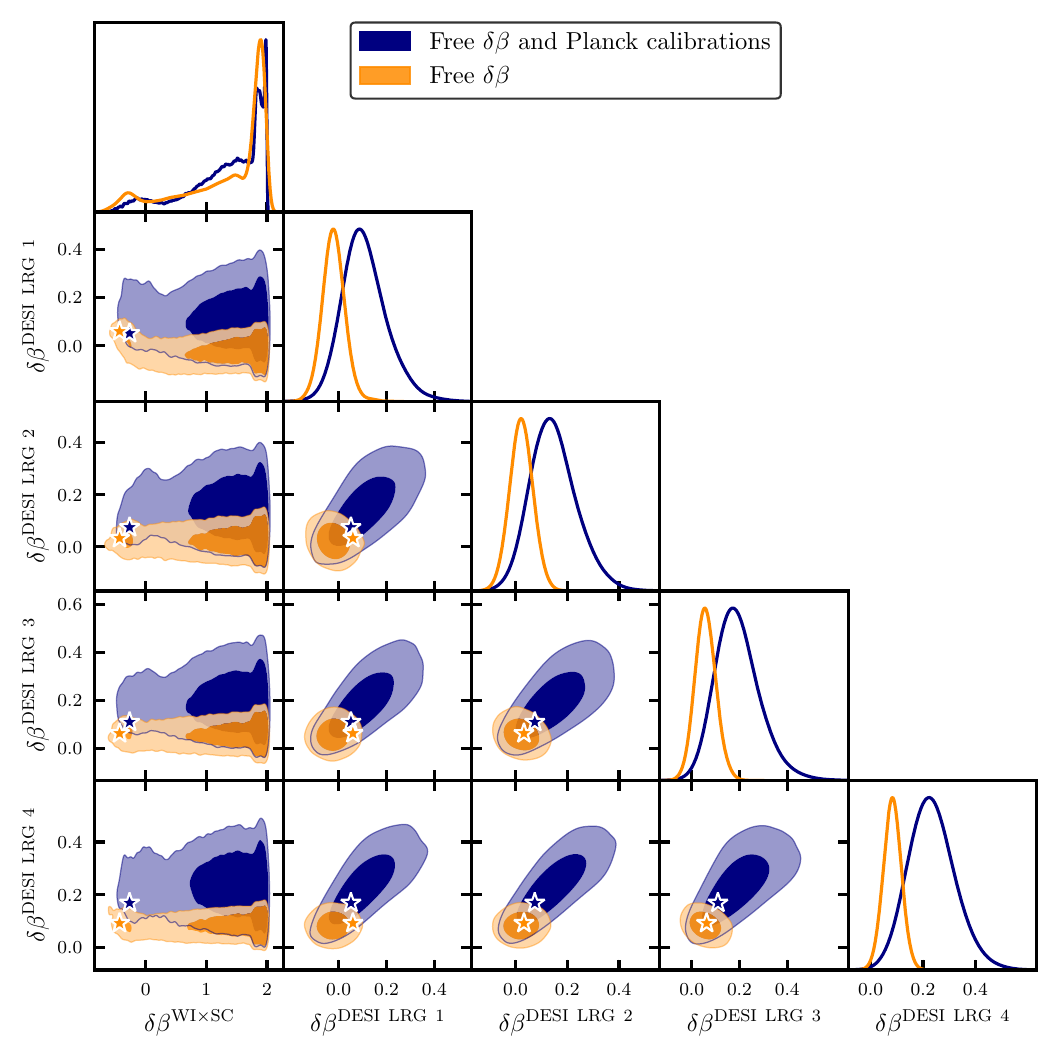}
      \caption{Marginalized posterior distribution on the $\delta\beta$ parameters, quantifying the departure from the CIB SED assumed in the analysis in each redshift bin. We find no significant deviation from the SEDs used in the baseline analysis ($\delta\beta=0$). We note strong volume effects affecting $\delta\beta^{\mathrm{WI}\times\mathrm{SC}}$ as the signal is consistent with zero in the redshift bin.}\label{fig:betas}
    \end{figure*}
    
    Galactic dust is present in addition to the CIB in all the \planck frequency channels, and can affect the observed galaxy number counts on large scales via extinction, thus constituting a potentially correlated systematic. This is of particular concern for the \wisc sample, which displays rather low values of both $\bpe$ and $\bsfr$. The test discussed above is somewhat sensitive to this effect, given the different frequency dependence of Galactic dust and the CIB. To further test for this effect, we repeated our analysis using the more conservative \planck Galactic mask covering 40\% of the sky (we used the 60\% mask in our fiducial analysis). We also tested the impact of removing the largest scales, where Galactic contamination is more prominent, from the analysis. Specifically, we considered the impact of changing our low-$\ell$ cut to $\ell_{\rm min}=50$ or $\ell_{\rm min}=80$. The results of these tests are shown in the right panel of Fig. \ref{fig:robustness_exp}. We do not find any significant deviation in the reconstructed $\bpe$ and $\bsfr$ compared to the statistical uncertainties.

    To further understand the low values reconstructed from the \wisc sample, we substituted this low redshift sample with the DECALS sample~\cite{2010.00466} and found consistent low values of both $\bpe$ and $\bsfr$. The same conclusion holds if we do not use the Planck point source mask for power spectrum estimation to quantify the impact of resolved sources.
    
    As pointed out in~\cite{2307.01043, 2307.01258}, \planck map-level calibration amplitude uncertainties can be quite large at high frequencies and particularly for the 857~GHz channel, which is still using a planet-based calibration in the \planck PR4 data release~\cite{2007.04997}. Unlike the MLBR or standard map-based ILC methods, it is relatively simple to marginalise over calibration amplitudes in our multi-frequency forward model. The square data points in the right panel of Fig. \ref{fig:robustness_exp} show the results after marginalising over the calibration amplitudes with a Gaussian priors based on uncertainties quoted in~\cite{2007.04997} and setting the standard deviation to $10\%$ at 857 GHz. We find that this marginalisation does not change or degrade our measurements significantly.

    Finally, as described in the main text, we considered the impact of variations in the CIB spectral tilt with respect to the fiducial SED model used here (modifying the spectrum by a factor $\propto\nu^{\delta\beta}$, with a different $\delta\beta$ in each redshift bin), finding that our measurements are largely robust to this potential source of uncertainty. Figure \ref{fig:betas} shows the posterior distribution of the spectral index variations $\delta\beta$ in each redshift bin. Constraints are shown for our fiducial analysis, in orange. We are able to constrain $\delta\beta$ at the level of $\Delta(\delta\beta)\sim0.1$ in all cases other than the \wisc redshift bin, where the CIB correlation is not detected (and hence we are not able to measure $\delta\beta$). In all cases we find that $\delta\beta$ is compatible with zero within $\sim1$-$2\sigma$, suggesting that our fiducial CIB model, based on the spectral templates of \citep{1304.3936,1409.5796,1703.08795}, provides a good representation of the observed CIB spectrum. The blue contours in the figure show the constraints found when simultaneously marginalising over $\delta \beta^g$ and the per-frequency calibration factors. Both parameters are mutually correlated, since they modify the overall amplitude of the CIB signal ($\delta\beta$ does so through the normalisation of $S_\nu^{\rm CIB}$). We find that the marginalised posterior of $\delta\beta^g$ broadens considerably, and is shifted to larger values.  We find, however, that the best-fit value of $\delta\beta^g$ does not change significantly with respect to those found for constant calibration factors, showing that this shift is largely a volume/projection effect caused by marginalising over a highly-degenerate set of parameters.

  \subsection{Correlated errors}
    \begin{figure}
      \centering
      \includegraphics[width=0.49\textwidth]{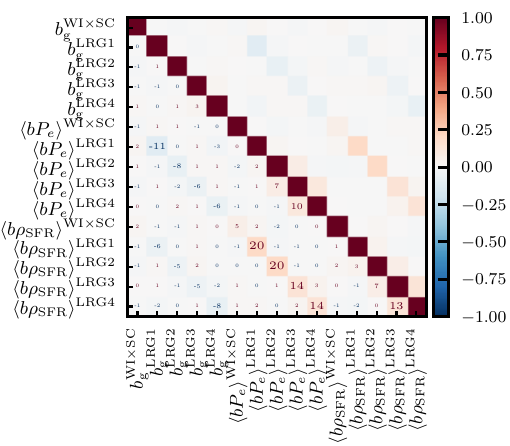}
      \includegraphics[width=0.49\textwidth]{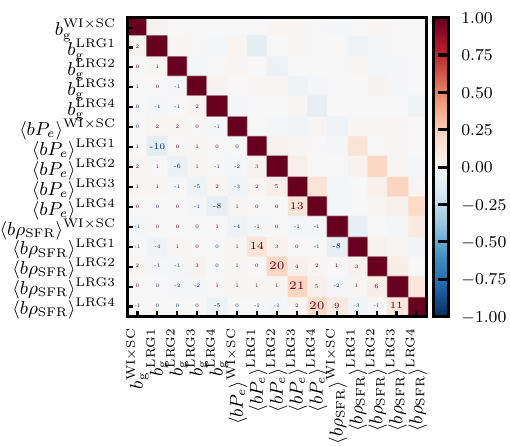}
      \caption{Parameter correlations in the baseline analysis (a) and when marginalizing over the shape of the CIB SEDs within each redshift bin. We see that the correlation between the reconstructed bias quantities remains low ($\sim 15-20\%$) in both cases.}\label{fig:corrs}
    \end{figure}
    The posterior distribution of the $\bpe$ and $\bsfr$ measurements presented here are very close to Gaussian \cite{2206.15394}, and Table I in the main text provides the mean and standard deviation of all parameters. The only remaining piece of information needed to incorporate these measurements as external data in any large-scale structure analysis is an estimated of their correlated errors. The left panel of Fig.~\ref{fig:corrs} shows the correlation matrix $\rho_{ij}\equiv {\rm Cov}_{ij}/\sqrt{{\rm Cov}_{ii}{\rm Cov}_{jj}}$ for our measurements of $b_g$, $\bpe$, and $\bsfr$ in each redshift bin for our fiducial analysis. Although the redshift distributions of the different samples used here overlap, the correlation between measurements in different bins is relatively small, with a maximum of $13\%$ for $\bsfr$ in the two highest redshift bins, which feature the broadest redshift tails. Within a single redshift bin, the measurements of $\bpe$ and $\bsfr$ show the strongest (although still mild) correlations of up to $\sim20\%$. The right panel of the same figure shows the correlation matrix when marginalising over the CIB SED tilts $\delta\beta^g$, which displays largely the same structure as our fiducial analysis. The numerical values of $\rho_{ij}$ for our fiducial analysis are:
    \begin{equation}
      \mathsf{\rho}=\left(
      \begin{array}{ccccc|ccccc}
        1.00 & 0.00 & -0.02 & -0.01 & -0.01 & 0.05 & -0.01 & -0.00 & 0.01 & 0.01 \\
        0.00 & 1.00 & 0.02 & 0.01 & -0.00 & 0.02 & 0.20 & -0.00 & -0.00 & 0.02 \\
        -0.02 & 0.02 & 1.00 & 0.07 & -0.01 & -0.02 & -0.01 & 0.20 & 0.01 & -0.00 \\
        -0.01 & 0.01 & 0.07 & 1.00 & 0.10 & -0.00 & -0.01 & -0.01 & 0.14 & 0.02 \\
        -0.01 & -0.00 & -0.01 & 0.10 & 1.00 & 0.00 & -0.00 & -0.00 & 0.03 & 0.14 \\ \hline
        0.05 & 0.02 & -0.02 & -0.00 & 0.00 & 1.00 & 0.01 & 0.02 & 0.00 & -0.01 \\
        -0.01 & 0.20 & -0.01 & -0.01 & -0.00 & 0.01 & 1.00 & 0.03 & -0.01 & -0.02 \\
        -0.00 & -0.00 & 0.20 & -0.01 & -0.00 & 0.02 & 0.03 & 1.00 & 0.07 & 0.00 \\
        0.01 & -0.00 & 0.01 & 0.14 & 0.03 & 0.00 & -0.01 & 0.07 & 1.00 & 0.13 \\
        0.01 & 0.02 & -0.00 & 0.02 & 0.14 & -0.01 & -0.02 & 0.00 & 0.13 & 1.00 \\
      \end{array}
      \right).
    \end{equation}
    
\end{document}